\documentclass[prb,preprint,superscriptaddress,nobibnotes]{revtex4}

\usepackage{times}
\usepackage[pdftex]{graphicx}
\usepackage[pdftex]{epsfig}
\usepackage{amsmath}
\usepackage{pslatex}
\usepackage{amssymb}
\usepackage{amsfonts}
\usepackage{color}
\usepackage{wrapfig}
\usepackage{eucal}
\usepackage{hhline}
\usepackage{threeparttable}
\usepackage{supertabular}
\usepackage{multirow}
\usepackage{tabularx}

\newcolumntype{Y}{>{\centering\arraybackslash}X}

\newcommand{\ave}[1]{\langle #1 \rangle}%

\begin{document}

\pagenumbering{arabic}

\title{Static and Dynamic Wavelength Routing via the Gradient Optical Force}

\author{Jessie Rosenberg}
\author{Qiang Lin}
\author{Kerry J. Vahala}
\author{Oskar Painter}
\email{opainter@caltech.edu}
\homepage{http://copilot.caltech.edu}
\affiliation{Thomas J. Watson, Sr., Laboratory of Applied Physics, California Institute of Technology, Pasadena, CA 91125}

\date{\today}

\begin{abstract}
Here we propose and demonstrate an all-optical wavelength-routing approach which uses a tuning mechanism based upon the optical gradient force in a specially-designed nano-optomechanical system.  The resulting mechanically-compliant ``spiderweb'' resonantor realizes seamless wavelength routing over a range of 3000 times the intrinsic channel width, with a tuning efficiency of 309~GHz/mW, a switching time of less than 200~ns, and 100\% channel-quality preservation over the entire tuning range. These results indicate the potential for radiation pressure actuated devices to be used in a variety of photonics applications, such as channel routing/switching, buffering, dispersion compensation, pulse trapping/release, and widely tunable lasers.
\end{abstract}

\maketitle


Optical information processing in photonic interconnects relies critically on the capability for wavelength management \cite{ref:Beausoleil, ref:Berthold}. The underlying essential functionalities are optical filtering and wavelength routing, which allow for precise selection and flexible switching of optical channels at high speeds over a broad bandwidth \cite{ref:Beausoleil, ref:Gripp, ref:Strand, ref:Neilson}. In the past two decades, a variety of technologies have been developed for this purpose \cite{ref:Sadot, ref:Eldada, ref:Tomlinson}; those based on micro/nano-resonators are particularly attractive because of their great potential for future on-chip integrated photonic applications \cite{ref:Little4, ref:Yi, ref:Rabiei, ref:Raineri, ref:Asano, ref:Poon, ref:Xu3, ref:Ilchenko, ref:Wu, ref:Barwicz, ref:Monat, ref:Xia, ref:Sherwood, ref:Vlasov, ref:Fushman}. In general, reconfigurable tuning of cavity resonances is realized through thermo-optic \cite{ref:Asano, ref:Sherwood, ref:Rabiei, ref:Fushman}, electro-optic \cite{ref:Xu3, ref:Raineri, ref:Vlasov, ref:Fushman}, photochemical \cite{ref:Poon}, optofluidic \cite{ref:Monat}, or microelectricalmechanical approaches \cite{ref:Yi,ref:Yano, ref:Wu}. However, all of these tuning mechanisms have intrinsic limitations on their tuning speed \cite{ref:Asano, ref:Rabiei, ref:Sherwood, ref:Poon, ref:Monat, ref:Yi, ref:Wu}, tuning bandwidth \cite{ref:Raineri, ref:Vlasov, ref:Xu3, ref:Fushman}, routing efficiency \cite{ref:Yi, ref:Rabiei, ref:Fushman, ref:Vlasov}, and/or routing quality \cite{ref:Xu3,ref:Fushman,ref:Vlasov,ref:Raineri}. Here we propose and demonstrate an all-optical wavelength-routing technique based upon actuation and tuning via radiation pressure forces.  This approach enables both wideband tuning and fast switching rates, with high channel-quality preservation over the entire tuning range.

The physics of electromagnetic forces within mechanically-compliant resonant cavities is by now well established, with some of the early experimental considerations being related to the quantum-limited measurement of weak, classical forces\cite{Braginsky77}.  In the optical domain, experiments involving optical Fabry-Perot ``pendulum cavities'' were first explored\cite{ref:Dorsel}, with more recent studies having measured radiation pressure forces in micro- and nano-mechanical cavities\cite{Kippenberg05,ref:Gigan06,Arcizet06,ref:Kleckner06,ref:Thompson3,ref:Li2,Eichenfield09}.  In each of these systems, whether it be gravitational wave observatory\cite{Abramovici92} or photonic crystal nanomechanical cavity\cite{Eichenfield09}, the same fundamental physics applies.  A narrowband laser input to the system, of fixed frequency, results in a ``dynamical back-action''\cite{Braginsky92} between mechanical fluctuations and the internal electromagnetic field.  This dynamical back-action modifies both the real and imaginary parts of the frequency of the mechanical motion, yielding an optically-controllable, dynamic mechanical susceptibility.  A seperate effect occurs when the laser frequency is swept across the cavity resonance, pushing on the mechanical system as the internal light field builds up near cavity resonance.  The more compliant the mechanical system, the larger the static displacement and the larger the tuning of the optical cavity.  Here we utilize both the static and dynamic mechanical susceptibilities of a coupled opto-mechanical system to realize a chip-based optical filter technology in which wideband tuning and fast switching can be simultaneously accomplished.                       

\begin{figure}[t]
 \begin{center}
 \includegraphics[width=0.8\columnwidth]{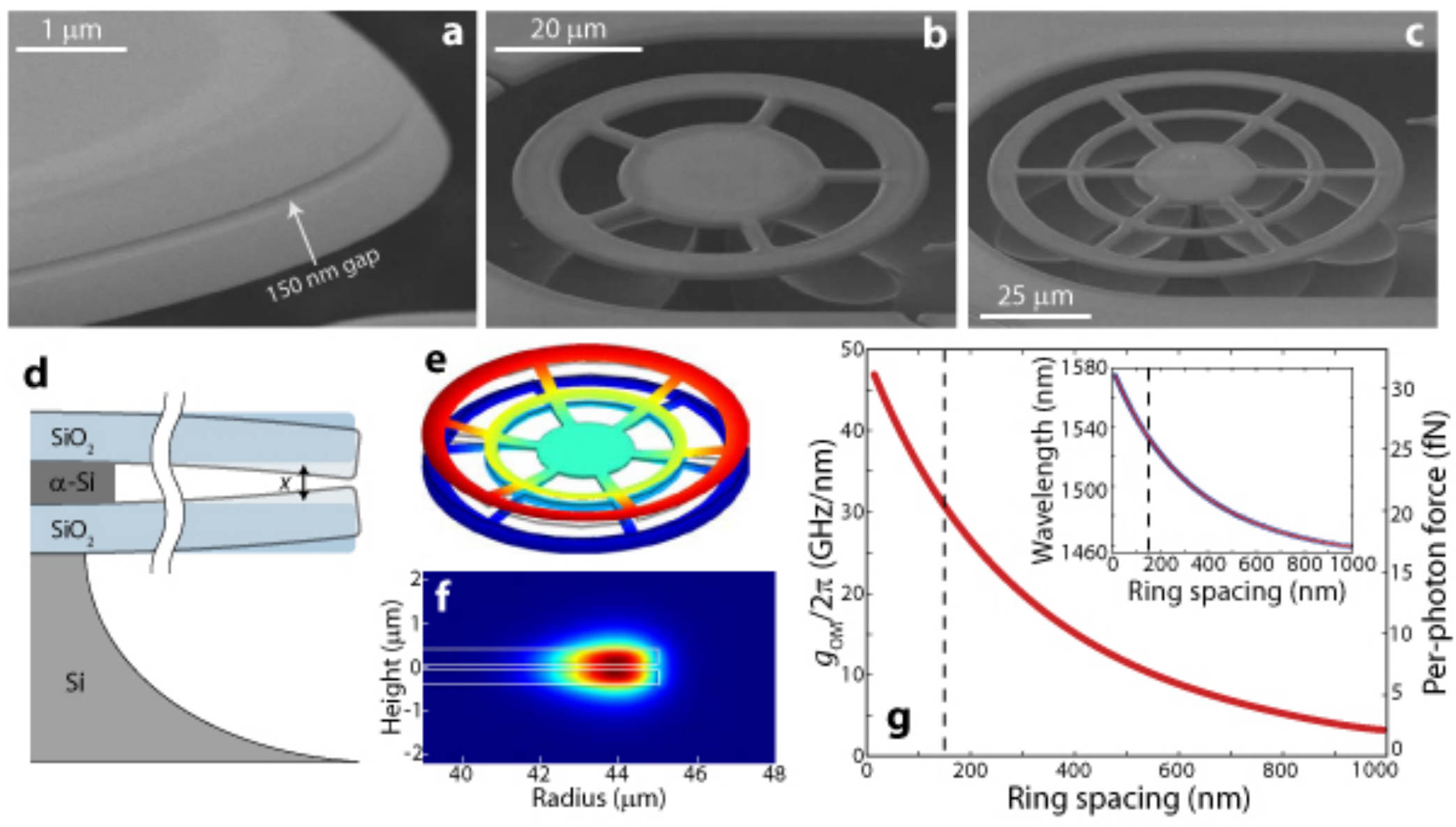}
 \caption{\textbf{Spiderweb microresonator images and simulations}. Scanning electron microscope images of \textbf{a}, the between-ring gap, \textbf{b}, the 54~$\mu$m spiderweb resonator, and \textbf{c}, the 90~$\mu$m spiderweb resonator. \textbf{d}, Schematic of a cross-section of the resonator, showing the bending of the two silica rings under the influence of the optical force. \textbf{e}, Mechanical FEM simulation of the bending of the 90~$\mu$m spiderweb resonator. The outward bending motion is shown for ease of viewing, and is exaggerated for clarity. \textbf{f}, FEM simulation of the radial component of the electric field for the fundamental TE bonding mode of the 90~$\mu$m spiderweb structure. \textbf{g}, The wavelength tunability, per-photon force and wavelength (inset) of the spiderweb cavity as a function of the ring spacing. The blue circles in the inset are from finite element simulation, and the inset red curve is a theoretical fit. The vertical dashed lines represent the experimentally-realized ring spacing of 150~nm.}
 \label{fig:Fig1}
 \end{center}
 \end{figure}

The optomechanical system we consider here is a simple modification to the common microring whispering-gallery cavity that has found widespread application in microphotonics.  As shown in Fig.~\ref{fig:Fig1}d, it consists of a pair of planar microrings, one stacked on top of the other.  The resulting near-field modal coupling forms a ``super-cavity,'', with a resonance frequency $\omega_0$ strongly dependent on the vertical cavity spacing, $x$.  The optomechanical coupling coefficient, $g_{\text{OM}} \equiv d\omega_0/dx$, determines both the tunability and per-photon optical force\cite{ref:Povinelli051, ref:Rakich}.  Finite-element-method (FEM) simulation shows that, for two 400-nm-thick planar silica whispering-gallery microcavities placed 150~nm apart (Fig.~\ref{fig:Fig1}f and g), the resonance tunability is as large as $g_{\text{OM}}/2\pi=31$ GHz/nm (corresponding to a $21$ fN/photon force).  The corresponding static mechanical displacement for $N$ photons stored inside the cavity is $\Delta x_{\text{static}} = {N \hbar g_{\rm OM}}/{k}$, where $k$ is the intrinsic spring constant of the mechanical structure. The overall magnitude of the cavity resonance tuning is then,

\begin{equation}
\Delta \omega_0 = g_{\text{OM}} \Delta x_{\text{static}} = \frac{N \hbar g_{\rm OM}^2}{k} = \frac{g_{\rm OM}^2 P_{\text{d}}}{k \omega_{0} \Gamma_0}, \label{D_omega0}
\end{equation}

\noindent where $P_{\text{d}}$ is the power dropped into the cavity and $\Gamma_0$ is the intrinsic photon decay rate, inversely proportional to the optical quality factor.

As the optical gradient force stems from the evanescent field coupling between the two near-field-spaced cavities, it is completely independent of the round-trip length of the cavity. This feature enables independent control of the optical and mechanical properties, allowing us to freely engineer the intrinsic mechanical rigidity through the scalability of the structure without changing the per-photon force.  In order to minimize the mechanical stiffness while also providing mechanical stability, we utilize a spiderweb-like support structure consisting of an arrangement of spokes and inner rings (see App. \ref{AppA} for fabrication details).  The zeroth-order spiderweb cavity (Fig.~\ref{fig:Fig1}b) has a 54~$\mu$m diameter outer ring supported by five spokes, while the first-order structure (Fig.~\ref{fig:Fig1}c) has a 90~$\mu$m outer diameter ring with six spokes and one supporting inner ring. FEM simulations show that these structures have spring constants of 9.25~N/m and 1.63~N/m for the smaller and larger resonators, respectively.

\begin{figure}[t]
 \begin{center}
 \includegraphics[width=0.75\columnwidth]{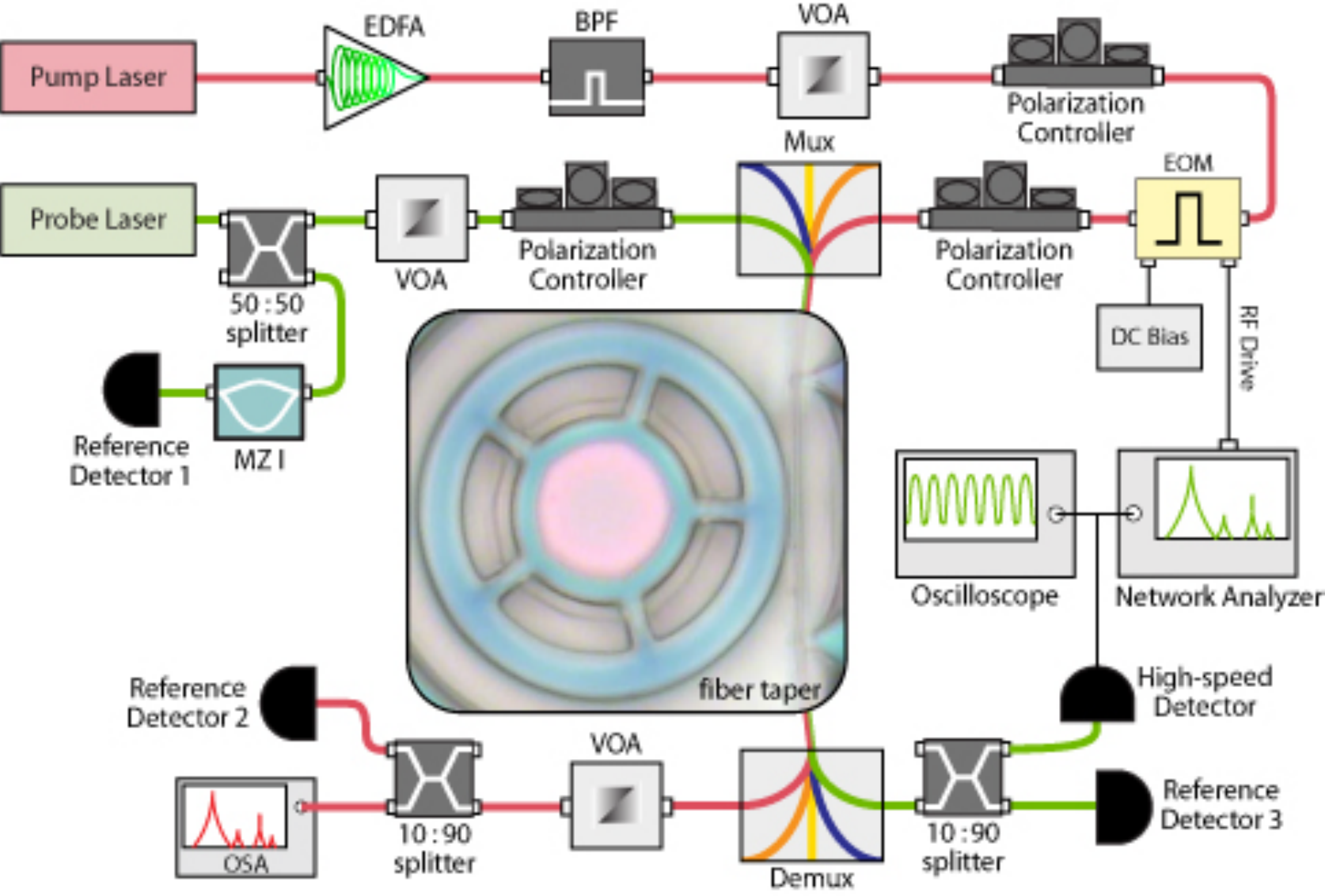}
 \caption{\textbf{Pump-probe experimental setup}. Schematic of the experimental setup for optical testing. The pump and probe lasers are coupled to the spiderweb resonator via a single-mode silica fiber taper stabilized by two nanoforks fabricated near the device. The pump laser power is boosted by an erbium-doped fiber amplifier (EDFA) and passed through a band-pass filter (BPF), and the two lasers are split into separate wavelength channels using a mux/demux system. For modulation experiments, the pump laser wavelength is modulated using an electro-optic modulator (EOM) driven by a network analyzer. The laser power levels are controlled by several variable optical attenuators (VOAs), the probe wavelength is calibrated by a Mach-Zehnder interferometer (MZI), and the pump wavelength is monitored by an optical spectrum analyzer (OSA). The spiderweb device itself is contained within a nitrogen environment at atmospheric pressure.}
 \label{fig:Fig2}
 \end{center}
 \end{figure}

\begin{figure}[t]
 \begin{center}
 \includegraphics[width=0.8\columnwidth]{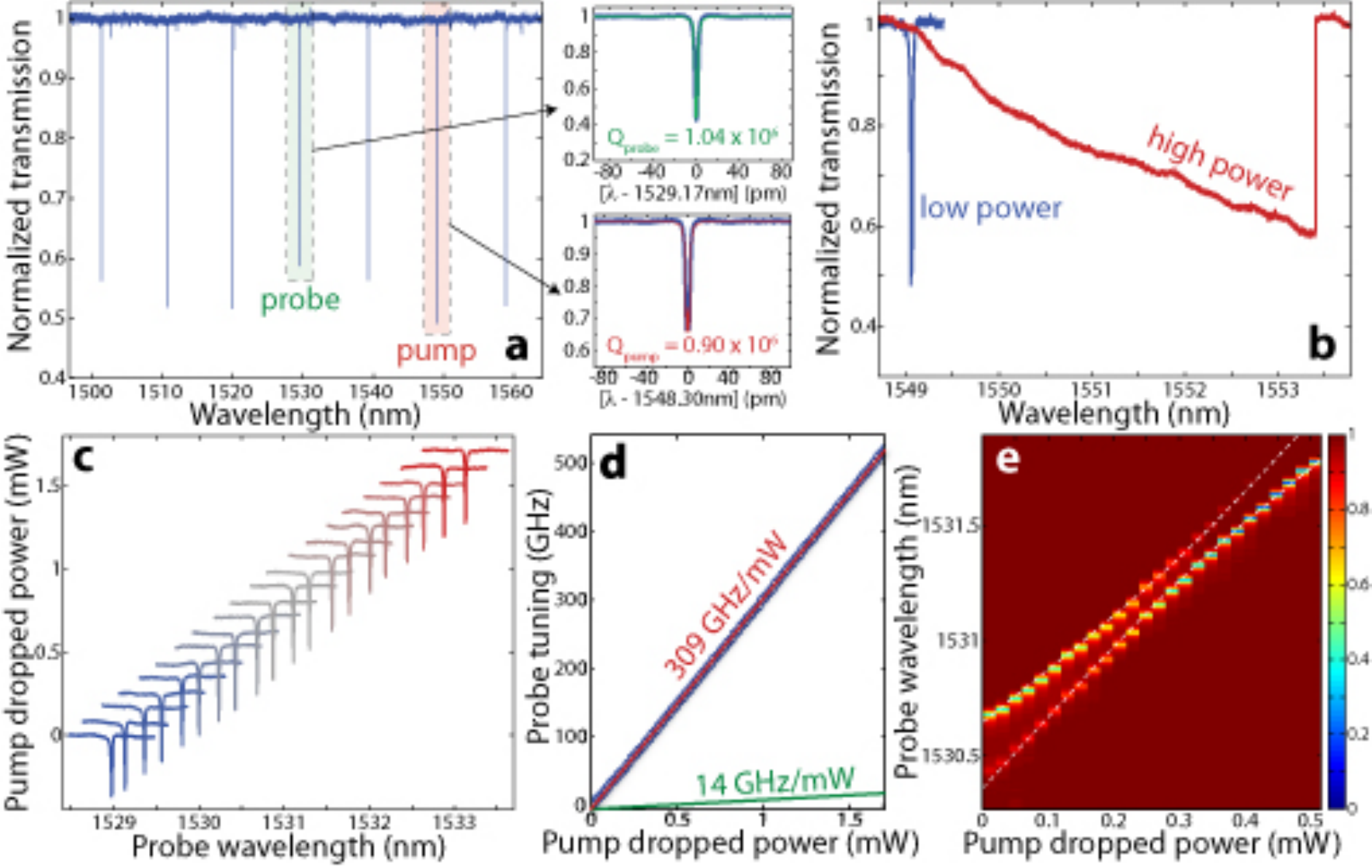}
 \caption{\textbf{Static tuning of a spiderweb microresonator}. \textbf{a}, Transmission spectrum of the 54~$\mu$m spiderweb cavity, recorded at the slow reference detector 3 (see Fig.~\ref{fig:Fig2}). The probe mode is shown highlighted in green, and the pump mode in red. Optical quality factor measurements of both modes are shown in the right-side insets, with theoretical fits shown in green and red for the pump and probe modes respectively. \textbf{b}, Scanned pump-mode transmission spectrum at 275~nW (blue) and 1.7~mW (red) dropped power. \textbf{c}, Probe-mode transmission curves for a selection of dropped powers in \textbf{d}, with the pump-mode dropped power indicated by the baseline of each transmission curve. \textbf{d}, Tuning of the probe mode with varying pump-mode dropped power. Blue circles mark the experimentally-measured probe wavelengths, and the red line is a fit to the data. The green curve represents the thermo-optic component of tuning. \textbf{e}, Transmission spectrum demonstrating the anticrossing of two secondary probe modes. The white dashed curves represent a theoretical fit.}
 \label{fig:Fig3}
 \end{center}
 \end{figure}

In adddition to the favorable mechanical properties, the whispering-gallery nature of the spiderweb resonator provides for high-$Q$ optical resonances.  Figure \ref{fig:Fig3}a shows the low power, in-plane polarized, wavelength scan of a 54-$\mu$m diameter resonator.  The excited family of resonances, corresponding to the fundamental transverse-electric-like (TE-like) modes, has a free-spectral range (FSR) of $9.7$~nm, with resonances at $\lambda=1529$~nm and $\lambda=1549$~nm exhibiting intrinsic quality factors of $Q=1.04\times 10^6$ and $Q=0.90 \times 10^6$, respectively. The combination of high cavity $Q$-factor, large $g_{\text{OM}}$, and floppy spiderweb structure result in the large optomechanical bistability shown in Fig.~\ref{fig:Fig3}b. With a power of $1.7$~mW dropped into the cavity, the cavity resonance initially at $\lambda=1549$~nm is shifted by $4.4$~nm (a little more than $0.5$ THz), corresponding to a static mechanical displacement of $\Delta x_{\rm static} = 17.7$~nm.  We observed similar performance from the larger 90~$\mu$m spiderweb structures, although device yield ($20\%$) and a slow change in device properties over time (despite devices being tested in a nitrogen environment to avoid water adsorption), indicate that further mechanical design optimization may be necessary for the larger structures.  By comparison, the smaller 54-$\mu$m diameter structures had near-$100\%$ yield and maintained their properties over the entire period of testing.

As the mechanical displacement is universally experienced by all double-ring cavity modes, the displacement actuated by one cavity mode can be used to control the wavelength routing of an entire mode family, indicating a great potential for broad waveband translation and switching in the wavelength-division multiplexing configuration. This is demonstrated in Fig.~\ref{fig:Fig3}d, where the mechanical displacement actuated by the ``pump'' mode at $\lambda=1549$~nm is used to control the wavelength of a ``probe'' mode initially located at $\lambda=1529$~nm. With increased dropped pump power, the probe wavelength is tuned linearly and continuously by $4.2$~nm, approximately $3000$ times the probe resonance intrinsic channel linewidth (or $500$ times the loaded linewidth). This factor is at least one order of magnitude larger than any other conventional approach previously reported\cite{ref:Yi, ref:Raineri, ref:Asano, ref:Poon, ref:Ilchenko, ref:Wu, ref:Monat, ref:Sherwood, ref:Xu3, ref:Fushman, ref:Vlasov, ref:Rabiei}. The tuning range shown in Fig.~\ref{fig:Fig3}d is about $43\%$ of the FSR.  In principle, it is possible to tune over the entire free-spectral range with a moderate dropped pump power of only $4$~mW.  Importantly, this wavelength-routing approach is purely dispersive in nature and completely preserves the channel quality during the wavelength routing process as can be clearly seen in Fig.~\ref{fig:Fig3}c.  This is in contrast to other tuning mechanisms such as the electro-optic approach via carrier injection \cite{ref:Raineri, ref:Vlasov, ref:Xu3, ref:Fushman}, in which the accompanying carrier absorption degrades the quality of the switched channel and thus limits the ultimate tuning bandwidth.

Fitting the data in Fig.~\ref{fig:Fig3}d gives a tuning efficiency of $309$~GHz/mW. This value agrees reasonably well with the theoretically predicted value of $393$~GHz/mW inferred from optical and mechanical FEM simulations, with the difference likely attributable to the uncertaintity in the Young's modulus of the annealed PECVD silica. Detailed experiments (App. \ref{AppB}) show that the thermo-optic effect contributes only a small component to the overall tuning rate ($13.8$~GHz/mW), as shown by the green curve in Fig.~\ref{fig:Fig3}d, and FEM simulations indicate a negligible thermo-mechanical component ($\sim 0.06\%$).  This, and properties of the dynamical response of the system (see below),  show that the wavelength routing is indeed a result of the optical gradient force.

In addition to the TE-like modes, the spiderweb double-ring resonator also supports a family of high-$Q$ transverse-magnetic-like (TM-like) modes with a FSR of $10$~nm. FEM simulations show that the per-photon force is slightly larger for the TM modes ($26.5$~fN/photon, or a $59\%$ larger tuning efficiency).  Figure \ref{fig:Fig3}e shows the mode hybridization between a pair of TE and TM-like modes (the slight angle in the outer sidewall of the two rings breaks the vertical symmetry, allowing for mode-mixing) induced by the optical force tuning of the two mode families.  This experimental observation agrees well with a simple theory (dashed curve in Fig.~\ref{fig:Fig3}e; App. \ref{AppC}) in which the TM modes have a tuning efficiency $42\%$ larger than that of the TE modes. Such a precisely tunable channel coupling may find applications in polarization switching/multiplexing/demultiplexing in optical signal processing, or carrier-sideband filtering in microwave photonics \cite{ref:Seeds}.

\begin{figure}[t]
 \begin{center}
 \includegraphics[width=0.8\columnwidth]{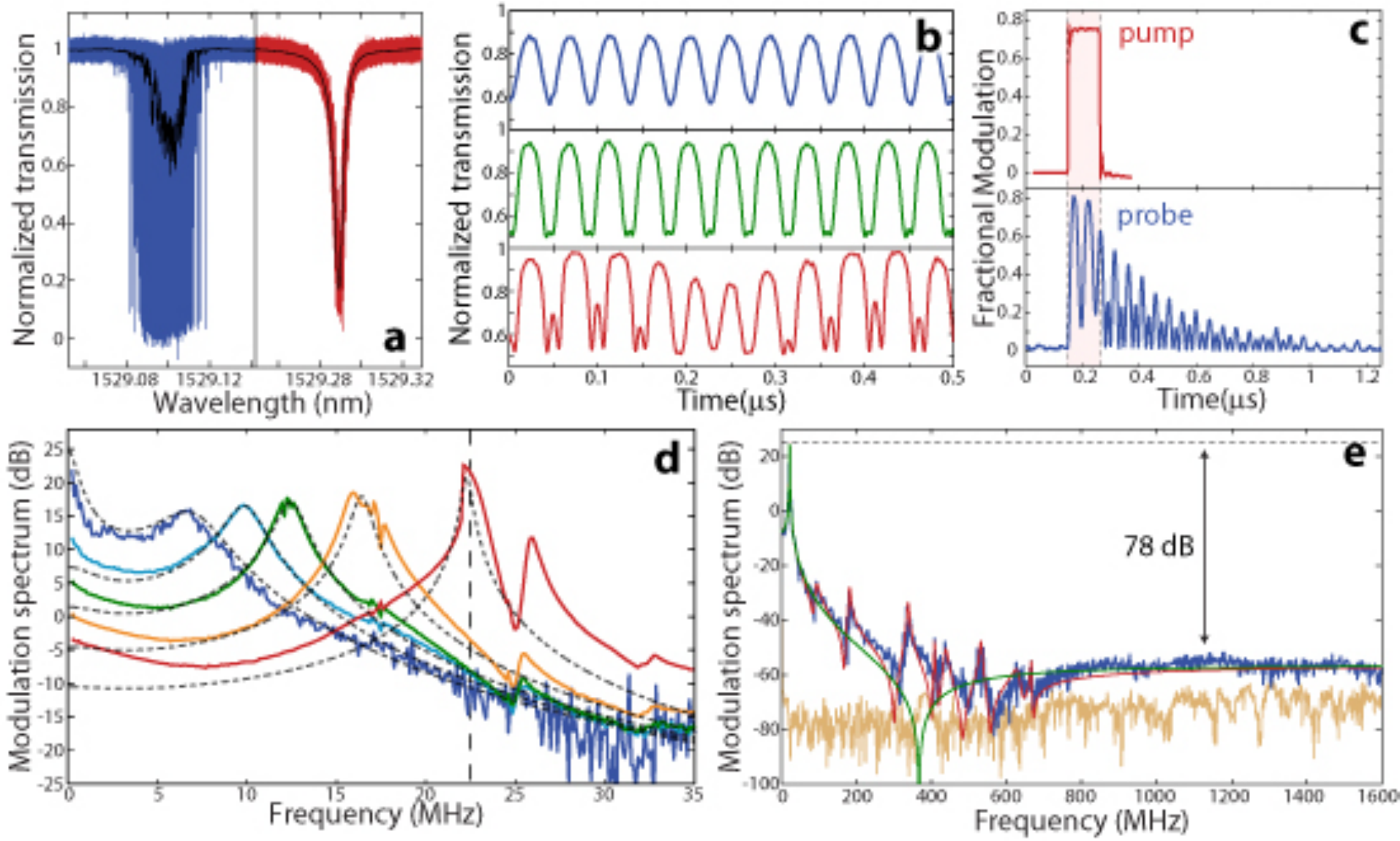}
 \caption{\textbf{Dynamic response of a spiderweb microresonator}. \textbf{a}, Probe transmission spectra recorded with the high-speed detector (Fig.~\ref{fig:Fig2}), at 0~mW (blue) and 0.20~mW (red) pump dropped power, with time-averaged traces shown in black.  \textbf{b}, Time waveforms of the probe transmission when the pump mode is sinusoidally modulated at 22.3~MHz (the vertical dashed line in \textbf{d}) with modulation depths of 1.9\% (blue), 14.9\% (green), and 20.5\% (red). \textbf{c}, Pulsed modulation of the pump (above) and corresponding probe response (below). The fractional modulation for the pump is defined as the relative modulation depth compared with the average dropped power, while that for the probe is the ratio to the full probe-mode coupling depth. \textbf{d}, Probe modulation spectra at pump dropped powers of 14~$\mu$W, 0.11~mW, 0.21~mW, 0.43~mW, and 0.85~mW (from left to right). For a better comparison of optical spring, the spectra are normalized by the dropped power ratio relative to the maximum 0.85~mW dropped power (see App. \ref{AppE}). The black dashed curves show the corresponding theory. \textbf{e}, Probe modulation spectrum at 0.85~mW pump dropped power, shown in a wide frequency span for comparison with the Kerr nonlinear response. The green curve shows the theory considering only the flapping mechanical mode, and the red curve shows the theory including multiple mechanical resonances. The orange curve represents the recorded detector noise floor.}
 \label{fig:Fig4}
 \end{center}
 \end{figure}

In addition to the static mechanical actuation of the spiderweb structure, as discussed above, the optical gradient force also introduces dynamical back action which alters the dynamic response of the mechanical motion \cite{Braginsky92,ref:Kippenberg06,ref:Favero2}.  The in-phase component of the optical force leads to a modified mechanical resonance frequency and effective dynamical spring constant of

\begin{equation}
\label{eq:spring_effect}
k' = k + \frac{2 g_{\text{OM}}^2 P_{\text{d}}\Delta }{\omega_0 \Gamma_0 [\Delta^2 + ({\Gamma_t}/{2})^2]},
\end{equation}

\noindent where $\Delta=\omega_{l}-\omega_{o}$ is the detuning of the input laser ($\omega_{l}$) from the cavity resonance ($\omega_{o}$) frequency and $\Gamma_t$ is the photon decay rate of the loaded cavity.  As the intrinsic spring constant of the spiderweb resonator is small ($9.5$ N/m), the dynamical spring can be greatly modified optically.  The alteration of the effective dynamic spring is clearly seen in the resonance spectra of the cavity resonances (left panel of Fig.~\ref{fig:Fig4}a).  For the floppy spiderweb structure, thermal Brownian motion introduces significant fluctuations in the cavity resonances (see App. \ref{AppD}).  As pump power is dropped into the cavity, however, the dynamic spring stiffens and strongly suppresses the magnitude of the thermal fluctuations (right panel of Fig.~\ref{fig:Fig4}a).

Optical control of the dynamic response is most clearly demonstrated through the pump-probe modulation response of the spiderweb structure.  Figure~\ref{fig:Fig4}d shows the spectral response of a probe resonance to small-signal sinusoidal pump modulation for several different (average) pump dropped powers. When the pump dropped power is low, the pump back-action on mechanical motion is negligible and the probe response is given by a combination of the intrinsic mechanical stiffness and the squeeze-film effect\cite{ref:Bao} of trapped gas in between the rings (App. \ref{AppE}). When the pump power is increased, however, the mechanical resonance frequency increases correspondingly, reaching a value of 22.3~MHz at a dropped power of 0.85~mW. This value is about 32 times larger than the intrinsic mechanical frequency, and implies a dynamical stiffness more than $1000$ times that of the silica rings.  The experimentally recorded probe modulation spectra can be described by the expression,

\begin{equation}
\rho(\Omega) = \left| {\frac{g_{\text{OM}}^2}{m_{\rm eff} \omega_p {\cal L}(\Omega) } + 2 \gamma } \right|^2 \frac{P_{\rm pd}^2}{\Gamma_{\rm 0p}^2} \frac{4 \Gamma_{\rm es}^2 \Gamma_{\rm 0s}^2 \Delta_s^2}{\left[ \Delta_s^2 + (\Gamma_{\rm ts}/2)^2 \right]^4}, \label{ProbeModulation}
\end{equation}

\noindent where $\gamma$ is the third-order Kerr nonlinear parameter and ${\cal L}(\Omega)$ is a spectral response including the intrinsic mechanical response, squeeze-film effect, and the dynamic back action from the optical gradient force (see App. \ref{AppE}). As shown in the dashed curves in Fig.~\ref{fig:Fig4}d, eq.~(\ref{ProbeModulation}) provides an accurate description of the pump-probe modulation response. Note that the Fano-like resonance at a power level of 0.85~mW is due to coherent mechanical mode mixing \cite{ref:Lin5}.

In general, the small modulation of the pump wave which actuates the mechanical oscillation is greatly magnified on the probe resonance.  Figure~\ref{fig:Fig4}d shows that for an average dropped pump power of only 0.85~mW there is a resonant modulation ``gain'' of greater than 20~dB.  This can also be seen in the time waveform of the probe in Fig.~\ref{fig:Fig4}b, where a 1.9\% modulation of the pump power is large enough to introduce considerable fractional modulation in the probe time waveform (top panel). Increasing the pump modulation to $14.9\%$ (Fig.~\ref{fig:Fig4}b, middle panel) results in a probe modulation of larger than a half-linewidth (full contrast modulation).  Further increase in the pump modulation depth actuates flapping mechanical motion so intense it begins to excite a second mechanical mode (the Fano-like feature in Fig.~\ref{fig:Fig4}d), resulting in a beat signal with a period of $0.36$~$\mu$s on the probe time waveform.

One metric for characterizing the response time of the spiderweb optomechanical cavity is the resonant oscillation period \cite{ref:Yano}. Figure \ref{fig:Fig4}d shows that the optical spring effect enables a modulation time as fast as 44.8~ns.  This can be further enhanced by using the transduction ``gain'' to push the probe modulation into the nonlinear regime, where in the lower panel of Fig.~\ref{fig:Fig4}b the probe wavelength (10\%--90\%) on-off switching time is reduced to 7~ns, roughly 3 orders of magnitude faster than modulation schemes based upon thermo-optic, optofluidic, photochemical, or microelectricalmechancial approaches \cite{ref:Yi, ref:Asano, ref:Poon, ref:Ilchenko, ref:Wu, ref:Monat, ref:Sherwood, ref:Rabiei,ref:Yano}.  For many switching applications, however, one is more interested in the impulse response of the system.  The pulsed response of the probe is shown in Fig.~\ref{fig:Fig4}c. As is common in micro/nanomechanical systems \cite{ref:Yano, ref:Wu}, the resonant response causes ringing during switching, with a settling time determined by the mechanical linewidth. The measured settling time of the probe response is $196$-ns, consistent with the mechanical linewidth of $\sim2-3$~MHz (see Fig.~\ref{fig:Fig4}d).

In addition to the optomechanical nonlinearity, other optical (material, etc.) nonlinearities can also contribute to the probe modulation. As shown in the expanded modulation spectrum of Fig.~\ref{fig:Fig4}e, the resonant optomechanical nonlinearity is dominant out to a frequency of $500$ MHz, after which the response plateaus due to the ultrafast Kerr nonlinearity of silica.  The Kerr nonlinearity is measured to be 78~dB below the resonant optomechanical response. This ratio agrees well with the theoretical value of 81~dB given by $\left(\frac{g_{\text{OM}}^2}{2 \gamma m_{\rm eff} \omega_p \Omega_m' \Gamma_m'}\right)^2$, where $\Omega_m'$ and $\Gamma_m'$ are the effective mechanical resonance frequency and damping rate, respectively (see App. \ref{AppE} for details).  The Kerr nonlinearity in silica has been extensively studied over more than three decades for optical signal processing \cite{ref:Agrawal2, ref:DelHaye, ref:Ferrera}, and the excellent agreement between the theoretical and experimental spectra provides yet another indication that the optical gradient force is the dominant tuning mechanism in the spiderweb cavity structure. 

The versatility of the gradient optical force tuning approach described here provides considerable room for future improvement of device performance.  An increase in the tuning range and efficiency (actuation power) can be expected with further engineering of the mechanical stability of the spiderweb structure.  For example, the 90 $\mu$m diameter first-order spiderweb cavities should allow for a six-fold increase in tuning efficiency to approximately $15$ nm/mW.  There are also many well-established methods for managing the dynamical response, in particular the ringing, of resonant micro- and nano-mechanical systems\cite{ref:Yano, ref:Wu}.  In contrast to cavity-optomechanical applications such as cooling and amplification of mechanical motion\cite{ref:Kippenberg06,ref:Favero2}, a reduction in the mechanical $Q$-factor, which can be obtained through elevated gas pressure or incorporation of damping materials, is sought to improve the switching time.  Given the similarity of the double-ring spiderweb stucture to other more conventional planar microring technologies, one can also incorporate other chip-based optical components such as waveguides, lasers, and modulators to enable full control and functionality of the optomechanics. One example technology would be an on-chip reconfigurable optical add/drop multiplexer or wavelength selective switch/crossconnect, which could be accomplished by integrating an array of double-ring cavities into a parallel or cascaded configuration. In addition to the demonstrated wavelength routing, other prospective applications for optomechanical devices include tunable optical buffering \cite{ref:Fontaine}, dispersion compensation \cite{ref:Madsen}, tunable lasers \cite{ref:Huang4}, and nonlinear signal processing\cite{ref:Agrawal2}. 

\begin{acknowledgments}
Funding for this work was provided by a DARPA seedling effort managed by Prof. Henryk Temkin, and the DARPA Phaser program.
\end{acknowledgments}


\appendix

\section{``Spiderweb'' resonator fabrication}
\label{AppA}
Fabrication of the spiderweb whispering-gallery resonator began with initial deposition of the cavity layers.  The two silica web layers and the sandwiched amorphous silicon ($\alpha$-Si) layer were deposited on a (100) silicon substrate by plasma-enhanced chemical vapor deposition, with a thickness of $400 \pm 4$~nm and $150 \pm 3$~nm for the silica and $\alpha$-Si layers, respectively. The wafer was then thermally annealed in a nitrogen environment at a temperature of $T=1050$ K for 10 hours to drive out water and hydrogen in the film, improving the optical quality of the material. The spiderweb pattern was created using electron beam lithography followed by an optimized C$_4$F$_8$-SF$_6$ gas chemistry reactive ion etch. Release of the web structure was accomplished using a SF$_6$ chemical plasma etch which selectively ($30,000:1$) attacks the intermediate $\alpha$-Si layer and the underlying Si substrate, resulting in a uniform undercut region which extends radially inwards 4~${\rm \mu}$m on all boundaries, fully releasing the web.  Simultaneously, the underlying silicon support pedestal is formed. Two nanoforks were also fabricated near the double-disk resonator to mechanically stabilize and support the fiber taper during optical coupling; the geometry was optimized such that the forks introduce a total insertion loss of only $\sim$4\%.

\section{Thermo-optic and thermo-mechanical contributions}
\label{AppB}
The thermo-optical effect on the resonance tuning was calibrated by using another identical device on the same sample. To isolate the thermo-optic effect from the optomechanical effect, we caused the two rings to stick together, eliminating the mechanical motion. Testing was performed on a cavity mode at 1552~nm using exactly the same conditions as for the wavelength routing measurements in the main text. A power of 2.1~mW dropped into the cavity introduces a maximum resonance red tuning by 0.23~nm, corresponding to a tuning rate of 0.11~nm/mW (13.8~GHz/mW), about 4\% of the total tuning rate recorded experimentally.

\begin{figure}[t]
 \begin{center}
 \includegraphics[width=0.6\columnwidth]{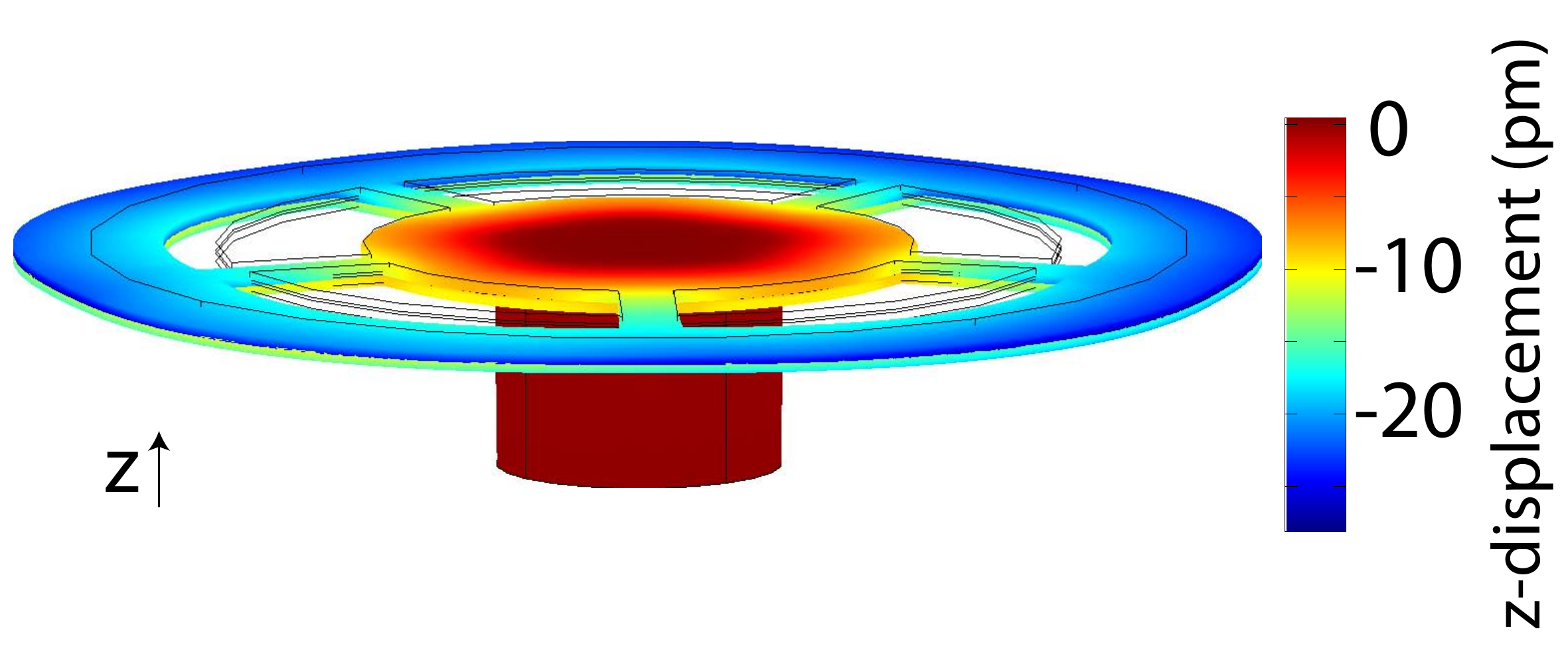}
 \caption{\textbf{Thermomechanical deflection of a spiderweb resonator}. FEM simulation illustrating the z-displacement of a 54~$\mu$m spiderweb resonator under the 21~K temperature differential between substrate and ring induced by 2.1~mW dropped optical power.}
 \label{fig:FigS1}
 \end{center}
 \end{figure}

The thermo-optic resonance tuning indicates a maximum temperature change of 21~K in the resonator. FEM simulations show that such a temperature variation of the resonator introduces a ring-gap change of only $\sim10$~pm, shown by the differential displacement of the top and bottom rings in Fig.~\ref{fig:FigS1}. Therefore, thermally induced mechanical deformation is expected to contribution a negligible 0.06\% of the experimentally recorded wavelength tuning.

The negligible contributions of both thermo-optic and thermo-mechanical effects are also confirmed by the pump-probe modulation spectra shown in Fig.~\ref{fig:Fig4}d-e, in which the magnitude and phase of the mechanical response relative to the Kerr response are consistent with optical gradient force actuation.  The thermo-mechanical effect, for instance, would be expected to be an additional $180$-degrees phase-lagged relative to the optical force, and thus would not show the dip in the pump-probe response upon interfering with the Kerr response (rather the two effects would be nearly in-phase with each other). 

\section{Anti-crossing of two probe modes}
\label{AppC}
The anti-crossing between the two probe modes when they approache each other is primarily due to the internal coupling between the two cavity modes, which can be described by a simple theory as follows. Assume two cavity resonances located at $\omega_{01}$ and $\omega_{02}$. For an input probe wave at $\omega$, the two cavity modes are excited through the following equations:
\begin{eqnarray}
\frac{da_1}{dt} &=& (i \Delta_1 - \frac{\Gamma_{\rm t1}}{2}) a_1 + i \beta a_2 + i \sqrt{\Gamma_{\rm e1}} A_{\rm in}, \label{da1_dt}\\
\frac{da_2}{dt} &=& (i \Delta_2 - \frac{\Gamma_{\rm t2}}{2}) a_2 + i \beta a_1 + i \sqrt{\Gamma_{\rm e2}} A_{\rm in}, \label{da2_dt}
\end{eqnarray}
where $\Delta_j = \omega - \omega_{\rm 0j}$ represents the cavity detuning of the $j^{\rm th}$ mode, and $\beta$ is the optical coupling coefficient between the two cavity modes. With a continuous-wave input, the steady state of Eqs.~(\ref{da1_dt}) and (\ref{da2_dt}) is given by the following solution
\begin{eqnarray}
a_1 &=& \frac{- i A_{\rm in} \left[ (i \Delta_2 - \Gamma_{\rm t2}/2) \sqrt{\Gamma_{\rm e1}} - i \beta \sqrt{\Gamma_{\rm e2}} \right]}{(i \Delta_1 - \Gamma_{\rm t1}/2)(i \Delta_2 - \Gamma_{\rm t2}/2) + \beta^2}, \label{a1}\\
a_2 &=& \frac{- i A_{\rm in} \left[ (i \Delta_1 - \Gamma_{\rm t1}/2) \sqrt{\Gamma_{\rm e2}} - i \beta \sqrt{\Gamma_{\rm e1}} \right]}{(i \Delta_1 - \Gamma_{\rm t1}/2)(i \Delta_2 - \Gamma_{\rm t2}/2) + \beta^2}. \label{a2}
\end{eqnarray}
As the transmitted field from the cavity is given by $A_T = A_{\rm in} + i \sqrt{\Gamma_{\rm e1}} a_1 + i \sqrt{\Gamma_{\rm e2}} a_2$, the cavity transmission thus has the following equation
\begin{equation}
T \equiv \frac{|A_T|^2}{|A_{\rm in}|^2} = \left| {\frac{(i \Delta_1 - \frac{\Gamma_{\rm 01} - \Gamma_{\rm e1}}{2})(i \Delta_2 - \frac{\Gamma_{\rm 02} - \Gamma_{\rm e2}}{2}) + (\beta - i \sqrt{\Gamma_{\rm e1} \Gamma_{\rm e2}} )^2} {(i \Delta_1 - \Gamma_{\rm t1}/2) (i \Delta_2 - \Gamma_{\rm t2}/2) + \beta^2 } } \right|^2. \label{T}
\end{equation}

\section{Quantifying the optical quality factor}
\label{AppD}
The extremely small intrinsic spring constant of the spiderweb resonator leads to significant thermal Brownian mechanical motion at low pump power (where the dynamic spring is small), and introduces considerable fluctuations on the cavity transmission spectrum (see Fig.~\ref{fig:Fig4}c). This makes it difficult to estimate the optical $Q$-factor of a cavity resonance at low pump power. As discussed previously, the thermal Brownian mechanical motion can be significantly suppressed through the optical spring effect. This feature provides an elegant way to accurately characterize the optical $Q$ of a cavity mode, by launching a relatively intense wave at a different resonance to suppress the perturbations induced by the thermal mechanical motion on a probe resonance. Moreover, a complete theory developed previously \cite{ref:Lin8} was used to describe the linear cavity transmission with the inclusion of the optomechanical effect.

\section{The cavity response in the pump-probe scheme}
\label{AppE}
Here we provide the derivation of the cavity response in a pump-probe scheme where the mechanical motion is primarily actuated by an intense pump and is experienced by a weak probe at a different resonance frequency.

\subsection{Pump field modulation}
In general, the pump wave inside the cavity satisfies the following equation:
\begin{equation}
\frac{da_p}{dt} = (i \Delta_p - \frac{\Gamma_{\rm tp}}{2}) a_p - i g_{\text{OM}}x a_p + i \gamma |a_p|^2 a_p + i \sqrt{\Gamma_{\rm ep}} A_{\rm p}, \label{dap_dt}
\end{equation}
where $a_p$ and $A_p$ are the intracavity and input field of the pump wave, respectively, normalized such that $U_p \equiv |a_p|^2$ and $P_p \equiv |A_p|^2$ represent the intracavity energy and input power. $\Delta_p = \omega_p - \omega_{\rm 0p}$ represents the detuning of pump frequency $\omega_p$ to the cavity resonance $\omega_{\rm 0p}$ and $\Gamma_{\rm tp}$ is the photon decay rate of the loaded cavity for the pump mode. In Eq.~(\ref{dap_dt}), the third term represents the back action of mechanical motion on the cavity resonance, where $g_{\text{OM}}$ is the optomechanical coupling coefficient and $x(t)$ is the mechanical displacement of the cavity structure. The fourth term describes the self-phase modulation introduced by the Kerr nonlinearity, where the nonlinear parameter $\gamma = \frac{ c \omega_p n_2}{n_0^2 V_{\rm eff}}$, $n_2=2.6 \times 10^{-20}~{\rm m^2/W}$ is the Kerr nonlinear coefficient of silica, $n_0=1.44$ is the silica refractive index, and $V_{\rm eff} = 370~{\rm \mu m^2}$ (from FEM simulation) is the effective mode volume \cite{ref:Agrawal2, ref:Lin6, ref:Lin7}. However, compared with the dominant optomechanical effect, the self-phase modulation on the pump wave is negligible in the spiderweb ring resonator. The final term in Eq.~(\ref{dap_dt}) represents the external field coupling with a photon escape rate of $\Gamma_{\rm ep}$.

Assume that the input pump wave consists of an intense continuous wave together with a small time-varying modulation, $A_p = A_{\rm p0} + \delta A_p(t)$. The intracavity field can be written as $a_p = a_{\rm p0} + \delta a_p(t)$, governed by the following equations:
\begin{eqnarray}
\frac{da_{\rm p0}}{dt} &=& (i \Delta_p - \frac{\Gamma_{\rm tp}}{2}) a_{\rm p0} + i \sqrt{\Gamma_{\rm ep}} A_{\rm p0}, \label{dap0_dt}\\
\frac{d \delta a_p}{dt} &=& (i \Delta_p - \frac{\Gamma_{\rm tp}}{2}) \delta a_p - i g_{\text{OM}}x a_{\rm p0} + i \sqrt{\Gamma_{\rm ep}} \delta A_{\rm p}, \label{dDap_dt}
\end{eqnarray}
where we have neglected the negligible self-phase modulation for the pump wave. Equation~(\ref{dap0_dt}) provides a steady-state solution of
\begin{equation}
a_{\rm p0} = \frac{i \sqrt{\Gamma_{\rm ep}} A_{\rm p0}}{\Gamma_{\rm tp}/2 - i \Delta_{p}}, \label{a_p0}
\end{equation}
from which we obtain the average pump power dropped into the cavity, $P_{\rm pd}$, given by
\begin{equation}
P_{\rm pd} = \frac{P_{\rm p0} \Gamma_{\rm 0p} \Gamma_{\rm ep}}{\Delta_p^2 + (\Gamma_{\rm tp}/2)^2}, \label{Pd}
\end{equation}
where $P_{\rm p0} = |A_{\rm p0}|^2$ is the averaged input pump power and $\Gamma_{\rm 0p}$ is the intrinsic photon decay rate of the pump mode. Clearly, to the zeroth order, the relative magnitude of the dropped pump power modulation is directly equal to that of the input modulation:
\begin{equation}
\frac{\delta P_{\rm pd}(t)}{P_{\rm pd}} = \frac{\delta P_p(t)}{P_{\rm p0}}, \label{dP_pd_Ratio}
\end{equation}
where $\delta P_p = A_{\rm p0}^* \delta A_p + A_{\rm p0} \delta A_p^* $ is the time-varying component of the input pump power.

Eq.~(\ref{dDap_dt}) leads to a pump-field modulation in the frequency domain of
\begin{equation}
\delta \widetilde{a}_p (\Omega) = \frac{i g_{\text{OM}} a_{\rm p0} \widetilde{x}(\Omega) - i \sqrt{\Gamma_{\rm ep}} \delta \widetilde{A}_p(\Omega)}{i(\Delta_p + \Omega) - \Gamma_{\rm tp}/2}, \label{da_p_Omega}
\end{equation}
where $\delta \widetilde{a}_p(\Omega)$, $\widetilde{x}(\Omega)$, and $\delta \widetilde{A}_p(\Omega)$ are Fourier transforms of $\delta {a}_p(t)$, ${x}(t)$, and $\delta {A}_p(t)$, respectively, defined as $\widetilde{B}(\Omega) = \int_{-\infty}^{+\infty} { B(t) e^{i\Omega t} dt}$. Physically, the first term in Eq.~(\ref{da_p_Omega}) represents the perturbation induced by the mechanical motion, while the second term represents the effect of direct input modulation.

\subsection{Optical gradient force}
The optical gradient force is linearly proportional to the cavity energy as $F_o = - \frac{g_{\text{OM}} U_p}{\omega_p}$. With modulation of the pump energy as discussed in the previous section, the gradient force thus consists of two terms, $F_o = F_{o0} + \delta F_o(t)$, where $F_{o0} = -\frac{g_{\text{OM}} U_{\rm p0}}{\omega_p}$ is the static force component introduced by the averaged pump energy $U_{\rm p0} = |a_{\rm p0}|^2$, and  $\delta F_o(t)$ is the dynamic component related to the pump energy modulation $\delta U_p(t)$, given by
\begin{equation}
\delta F_o(t) = -\frac{g_{\text{OM}} \delta U_p}{\omega_p} = -\frac{g_{\text{OM}}}{\omega_p} \left[a_{\rm p0}^* \delta a_p(t) + a_{\rm p0} \delta a_p^*(t)\right]. \label{DF_0}
\end{equation}
Substituting Eq.~(\ref{da_p_Omega}) into Eq.~(\ref{DF_0}), we find the force modulation is described by this general form in the frequency domain:
\begin{equation}
\delta \widetilde{F}_0(\Omega) = f_o(\Omega) \widetilde{x}(\Omega) + \frac{i\sqrt{\Gamma_{\rm ep}} g_{\text{OM}}}{\omega_p} \left[ \frac{a_{\rm p0}^* \delta \widetilde{A}_p(\Omega)}{i(\Delta_p + \Omega) - \Gamma_{\rm tp}/2} + \frac{a_{\rm p0} \delta \widetilde{A}_p^*(-\Omega)}{i(\Delta_p - \Omega) + \Gamma_{\rm tp}/2} \right], \label{DF_0_Omega}
\end{equation}
where the first term represents the back action introduced by the mechanical motion, with a spectral response $f_0(\Omega)$ given by
\begin{equation}
f_o(\Omega) \equiv - \frac{2 g_{\text{OM}}^2 |a_{\rm p0}|^2 \Delta_p }{\omega_p} \frac{\Delta_p^2 - \Omega^2 + (\Gamma_{\rm tp}/2)^2 + i \Gamma_{\rm tp} \Omega}{\left[(\Delta_p+\Omega)^2+ (\Gamma_{\rm tp}/2)^2 \right] \left[(\Delta_p - \Omega)^2 + (\Gamma_{\rm tp}/2)^2 \right]}. \label{f0_Omega}
\end{equation}

\subsection{The squeeze-film effect}

The spiderweb ring resonators are separated by a 150-nm gap, which is only about 2.2 times the mean free path in a nitrogen environment ($\sim$68~nm). As the ring is $\sim 6.3~\mu$m wide, much larger than the ring gap, the nitrogen gas sandwiched in the gap is highly confined by the two silica layers and cannot move freely during the flapping motion of the two rings. The resulting significant pressure differential between the internal and external regions of the paired silica rings functions as a viscous force to damp the mechanical motion. This phenomenon is well-known as the squeeze-film effect, which has a profound impact on the dynamic response of micro/nano-mechanical systems \cite{ref:Bao}. Apart from the optical gradient force, the squeeze film effect is the dominant mechanism responsible for the dynamic mechanical response of our devices. The associated damping force can be described by a general form of $\widetilde{F}_{\rm sq}(\Omega) = f_{\rm sq}(\Omega) \widetilde{x}(\Omega)$, where $f_{\rm sq}(\Omega)$ represents the spectral response of the squeeze film.

In general, the squeeze film effect is typically described by two theories which work in quite different regimes, depending on the Knudsen number $K_n$ characterizing the ratio between the mean-free path and the gap \cite{ref:Bao}. In the classical regime with $K_n \ll 1$ where the gas can be considered a continuum, the squeeze-film viscous force for a rectangular plate is well described by $f_{\rm sq}(\Omega) = - k_{e}(\Omega) + i C_d(\Omega)$, where $k_e$ and $C_d$ represents the spring constant and damping, respectively, induced by the squeeze film. They are given by the following equations \cite{ref:Blech}
\begin{eqnarray}
k_e (\Omega) &=& \frac{64 \sigma^2 P_a L_0 W_0}{\pi^8 h_0} \sum_{m,n~odd} {\frac{1}{m^2 n^2 \left[ (m^2 + (n/\eta)^2)^2 + \sigma^2/\pi^4\right]}}, \label{k_e}\\
C_d (\Omega) &=& \frac{64 \sigma P_a L_0 W_0}{\pi^6 h_0} \sum_{m,n~odd} {\frac{m^2+(n/\eta)^2}{m^2 n^2 \left[ (m^2 + (n/\eta)^2)^2 + \sigma^2/\pi^4\right]}}, \label{C_d}
\end{eqnarray}
where $P_a$ is the ambient gas pressure, $W_0$ and $L_0$ are the width and length of the plate, $h_0$ is the gap, $\eta = L_0/W_0$ is the aspect ratio of the plate, and $\sigma$ is the squeeze number given by
\begin{equation}
\sigma(\Omega) = \frac{12 \mu_{\rm eff} W_0^2 \Omega}{P_a h_0^2}, \label{sigma}
\end{equation}
where $\mu_{\rm eff} = \mu/(1+ 9.638 K_n^{1.159})$ is the effective value of the viscosity coefficient $\mu$ \cite{ref:Veijola}. Under this model, the squeeze film functions primarily as a damping (or elastic) force when the modulation frequency is below (or above) the cutoff frequency given by
\begin{equation}
\Omega_c = \frac{\pi^2 P_a h_0^2}{12 \mu_{\rm eff}} \left( \frac{1}{W_0^2} + \frac{1}{L_0^2} \right). \label{Omega_c}
\end{equation}
In contrast, in the free-molecule regime with $K_n \gg 1$ where the interaction between gas molecules is negligible, the squeeze film approximately behaves like a damping force, $f_{\rm sq}(\Omega) = i C_r \Omega$, with $C_r$ given by the following equation \cite{ref:Christian,ref:Bao2}
\begin{equation}
C_r = \left(\frac{S}{16 \pi h_0} \right) 4 P_a L_0 W_0 \sqrt{\frac{2 M_m}{\pi {\cal R} T}}, \label{C_r}
\end{equation}
where $M_m$ is the molar mass of gas, $T$ is the temperature, ${\cal R}$ is the ideal gas constant, $S$ is the perimeter length of the gap region.

However, our devices have a Knudsen number of $K_n = 0.45$, falling in the crossover regime where neither theory adequately describes the squeeze-film effect \cite{ref:Bhiladvala}. As the device works in the regime between the continuum and free-molecule limit, we heuristically propose that the damping/elastic force of the squeeze film is effectively described by a composite of the two theories:
\begin{equation}
f_{\rm sq}(\Omega) = - k_{e}(\Omega) + i C_d(\Omega) + i \eta_r C_r \Omega, \label{f_sq_Omega}
\end{equation}
with a modified effective coefficient of viscosity $\mu_{\rm eff}' = \eta_{\mu} \mu_{\rm eff}$, where $\eta_r$ and $\eta_\mu$ are parameters used for a best description of the squeeze-film response in our devices. Detailed analysis shows that $\eta_\mu = 0.7$ and $\eta_r=0.03$ provides the best fit for our devices. As our devices have a spiderweb geometry, we approximate it with an equivalent rectangular shape with $W_0$ given by the ring width, $L_0$ given by the circumference at the ring center, and $S \approx 2L_0$. As shown in the main text, this model provides an accurate description of the squeeze-film effect in our devices.

Although the intrinsic mechanical frequency of the 54~$\mu$m spiderweb structure is 694~kHz (indicated by FEM simulation), Fig.~4(a) in the main text shows a minimum dynamic frequency response of 6~MHz, dominated by the squeeze-film damping. Interestingly, although squeeze-film damping is generally detrimental in other micro/nanomechanical systems \cite{ref:Bao, ref:Verbridge1}, it is beneficial in this case, as it helps to extend the modulation bandwidth for wavelength routing.

\subsection{Actuated mechanical motion}
With the optical gradient force and the squeeze-film damping force, the mechanical motion of the cavity satisfies the following equation:
\begin{equation}
\frac{d^2 x}{dt^2} + \Gamma_m \frac{dx}{dt} + \Omega_m^2 x = \frac{1}{m_{\rm eff}}(F_o + F_{\rm sq} + F_T)  = \frac{1}{m_{\rm eff}}(F_{o0} + \delta F_o +F_{\rm sq} + F_T) , \label{dx_dt}
\end{equation}
where $m_{\rm eff}$ is the effective motional mass of the flapping mechanical mode, and $\Omega_m$ and $\Gamma_m$ are intrinsic mechanical frequency and damping rate, respectively. $F_T$ is the thermal Langevin force responsible for the thermal Brownian motion, a Markovin process with the following correlation function:
\begin{equation}
\ave{F_T(t) F_T(t+\tau)} = 2 m_{\rm eff} \Gamma_m k_B T \delta (\tau), \label{F_T_Corr}
\end{equation}
where $k_B$ is Boltzmann's constant.

As the squeeze-film viscous force is zero at $\Omega = 0$, the squeeze gas film impacts only the dynamic response of mechanical motion. Equation~(\ref{dx_dt}) shows clearly that the static mechanical displacement is actuated only by the static component of the optical force given by
\begin{equation}
x_0= \frac{|F_{o0}|}{m_{\rm eff} \Omega_m^2} = \frac{g_{\rm om} U_{\rm p0}}{k_m \omega_p} = \frac{g_{\rm om} P_{\rm pd}}{k_m \omega_p \Gamma_{\rm 0p}}, \label{x_0}
\end{equation}
where $k_m = m_{\rm eff} \Omega_m^2$ is the intrinsic spring constant of the spiderweb structure. With a specifically designed extremely small spring constant, $x_0$ can be quite significant for a given dropped power. As a result, the cavity resonance can be tuned by a significant magnitude of $g_{\text{OM}} x_0$. This is the primary mechanism responsible for the resonance tuning demonstrated in the main text. On the other hand, such a static mechanical displacement primarily changes the equilibrium position of the mechanical motion. It is convenient to remove this component in Eq.~(\ref{dx_dt}) by defining $x' = x - x_0$, since both the squeeze-film damping force and dynamic component of the optical force affect only the dynamics of $x'$.

Substituting Eqs.~(\ref{DF_0_Omega}), (\ref{f0_Omega}), (\ref{f_sq_Omega}) into Eq.~(\ref{dx_dt}) in the frequency domain, we find that the squeeze-film damping force and the backaction term of optical force primarily change the values of the resonant frequency and damping rate of the mechanical motions. Define
\begin{equation}
{\cal L}(\Omega) \equiv \Omega_m^2 - \Omega^2 - i \Gamma_m \Omega - \frac{f_o(\Omega)}{m_{\rm eff}} - \frac{f_{\rm sq}(\Omega)}{m_{\rm eff}}, \label{L_Omega}
\end{equation}
the mechanical displacement is thus given by
\begin{equation}
\widetilde{x}(\Omega) = \frac{\widetilde{F}_T(\Omega)}{m_{\rm eff} {\cal L}(\Omega)} + \frac{i\sqrt{\Gamma_{\rm ep}} g_{\text{OM}}}{m_{\rm eff} \omega_p {\cal L}(\Omega)} \left[ \frac{a_{\rm p0}^* \delta \widetilde{A}_p(\Omega)}{i(\Delta_p + \Omega) - \Gamma_{\rm tp}/2} + \frac{a_{\rm p0} \delta \widetilde{A}_p^*(-\Omega)}{i(\Delta_p - \Omega) + \Gamma_{\rm tp}/2} \right], \label{x_Omega}
\end{equation}
where we have dropped the prime notation of $x'$ for simplicity.

The first term in Eq.~(\ref{x_Omega}) represents the thermal Brownian motion while the second term describes the motions actuated by the pump modulation. In the absence of pump modulation, the mechanical motion is dominated by the Brownian motion. By using Eq.~(\ref{F_T_Corr}), we find the spectral density of thermal mechanical displacement has the form
\begin{equation}
{\cal S}_x(\Omega) = \frac{2 \Gamma_m k_B T}{m_{\rm eff}|{\cal L}(\Omega)|^2}. \label{S_x}
\end{equation}
Equations~(\ref{DF_0_Omega}), (\ref{f0_Omega}), and (\ref{L_Omega}) show that one dominant effect of the pump energy inside the cavity is to increase the mechanical rigidity, the so-called optical spring effect. In most cases, ${\cal L}(\Omega)$ can be well approximated by ${\cal L}(\Omega) \approx (\Omega_m')^2 - \Omega^2 - i \Gamma_m' \Omega$ with a new mechanical resonance $\Omega_m'$ and damping rate $\Gamma_m'$ affected by the optical force. Equation~(\ref{S_x}) thus leads to a variance of the thermal mechanical displacement given by
\begin{equation}
\langle (\delta x)^2 \rangle = \frac{1}{2 \pi} \int_{-\infty}^{+\infty}{{\cal S}_x(\Omega) d\Omega } = \frac{k_B T \Gamma_m}{k_m' \Gamma_m'} \approx \frac{k_B T}{k_m'}, \label{x_square}
\end{equation}
where $k_m' = m_{\rm eff} (\Omega_m')^2$ is the effective spring constant and the approximation in the final term assumes a negligible change in the mechanical linewidth. Clearly, the increase of the mechanical resonance frequency through the optical spring effect dramatically suppresses the magnitude of the thermal mechanical displacement and its perturbation of the cavity resonance, as shown clearly in the main text.

In the presence of pump modulation, the mechanical motion is primarily dominated by the dynamic optical force rather than the actuation from the thermal Langevin force, and the first term is negligible compared with the second term in Eq.~(\ref{x_Omega}). We focus on this situation in the following discussion, neglecting the thermal Brownian term.

\subsection{Cavity response of the probe mode}

The probe wave inside the cavity is governed by a dynamic equation similar to Eq.~(\ref{dap_dt}):
\begin{equation}
\frac{da_s}{dt} = (i \Delta_s - \frac{\Gamma_{\rm ts}}{2}) a_s - i g_{\text{OM}}x a_s + 2 i \gamma |a_p|^2 a_s + i \sqrt{\Gamma_{\rm es}} A_{\rm s}, \label{das_dt}
\end{equation}
except that the Kerr-nonlinear term now describes the cross-phase modulation from the pump wave. With the perturbations induced by the pump modulation, similar to the previous discussion of the pump wave, the intracavity probe field can be written as $a_s = a_{\rm s0} + \delta a_s(t)$, governed by the following equations:
\begin{eqnarray}
\frac{da_{\rm s0}}{dt} &=& (i \Delta_s - \frac{\Gamma_{\rm ts}}{2}) a_{\rm s0} + 2 i \gamma U_{\rm p0} a_{\rm s0} + i \sqrt{\Gamma_{\rm es}} A_{\rm s}, \label{das0_dt}\\
\frac{d \delta a_s}{dt} &=& (i \Delta_s - \frac{\Gamma_{\rm ts}}{2}) \delta a_s + 2 i \gamma U_{\rm p0} \delta a_{\rm s} - i g_{\text{OM}}x a_{\rm s0} + 2 i \gamma \delta U_p a_{\rm s0}, \label{dDas_dt}
\end{eqnarray}
where we have assumed the probe input is a continuous wave with a power of $P_s = |A_s|^2$. The second terms of Eqs.~(\ref{das0_dt}) and (\ref{dDas_dt}) represent the static cavity tuning introduced by cross-phase modulation, which can be included in the cavity tuning term $\Delta_s$ for simplicity. In general, it is negligible compared with the cavity linewidth at the power level used for exciting optomechanical effects, leading to $2\gamma U_{\rm p0} \ll \Gamma_{\rm tp}, \Gamma_{\rm ts}$.

Equation~(\ref{das0_dt}) provides a steady-state solution of
\begin{equation}
a_{\rm s0} = \frac{i \sqrt{\Gamma_{\rm es}} A_{\rm s}}{\Gamma_{\rm ts}/2 - i \Delta_{s}}, \label{a_s0}
\end{equation}
and Eq.~(\ref{dDas_dt}) results in a probe-field modulation in the frequency domain of
\begin{equation}
\delta \widetilde{a}_s (\Omega) = \frac{i a_{\rm s0} \left[ g_{\text{OM}} \widetilde{x}(\Omega) - 2 \gamma \delta \widetilde{U}_p(\Omega) \right] } {i(\Delta_s + \Omega) - \Gamma_{\rm ts}/2}, \label{da_s_Omega}
\end{equation}
where $\delta \widetilde{U}_p(\Omega)$ is the Fourier transform of $\delta U_p(t)$. As the transmitted field of the probe is given by $A_{\rm Ts} = A_s + i \sqrt{\Gamma_{\rm es}} a_s$, the modulation of the transmitted probe power thus takes the form
\begin{equation}
\delta P_{\rm Ts} = i \sqrt{\Gamma_{\rm es}} (A_{\rm 0s}^* \delta a_s - A_{\rm 0s} \delta a_s^*), \label{DP_t}
\end{equation}
where $A_{\rm 0s} = A_s + i \sqrt{\Gamma_{\rm es}} a_{\rm s0}$ is the transmitted probe wave in the absence of modulation. By use of Eqs.~(\ref{da_p_Omega}), (\ref{x_Omega}), (\ref{a_s0}) and (\ref{da_s_Omega}),  we find that the power spectrum of the transmitted probe modulation is given by the following equation:
\begin{equation}
\frac{|\delta \widetilde{P}_{\rm Ts}(\Omega)|^2}{P_s^2} = \left| {\frac{g_{\text{OM}}^2}{m_{\rm eff} \omega_p {\cal L}(\Omega) } + 2 \gamma } \right|^2 \frac{P_{\rm pd}^2}{\Gamma_{\rm 0p}^2} \frac{|\delta \widetilde{P}_{\rm pd}(\Omega)|^2}{P_{\rm pd}^2} \frac{4 \Gamma_{\rm es}^2 \Gamma_{\rm 0s}^2 \Delta_s^2}{\left[ \Delta_s^2 + (\Gamma_{\rm ts}/2)^2 \right]^4}, \label{ProbeModulation}
\end{equation}
where $\Gamma_{\rm 0s}$ is the intrinsic photon decay rate of the probe mode, and $\delta \widetilde{P}_{\rm Ts}(\Omega)$ and $\delta \widetilde{P}_{\rm pd}(\Omega)$ are the Fourier transforms of $\delta P_{\rm Ts}(t)$ and $\delta P_{\rm pd}(t)$, respectively. To obtain Eq.~(\ref{ProbeModulation}), we have used Eq.~(\ref{dP_pd_Ratio}) to relate the dropped pump power to the input, and have also taken into account the fact that the Kerr effect is relatively small, such that $2 \gamma U_{\rm p0} \ll \Gamma_{\rm tp}$. We also assume the cavity is in the sideband-unresolved regime with $\Omega_m \ll \Gamma_{\rm tp}, \Gamma_{\rm ts}$. The modulation spectra given in the main text are defined as
\begin{equation}
\rho(\Omega) \equiv \frac{|\delta \widetilde{P}_{\rm Ts}(\Omega)|^2 / P_s^2}{|\delta \widetilde{P}_{\rm pd}(\Omega)|^2 / P_{\rm pd}^2}. \label{Def_ModulationSpectra}
\end{equation}
For a better comparison of the dynamic-backaction induced variations on the probe modulation, the modulation spectra shown in Fig.~4d of the main text are normalized by a factor corresponding to the ratio of the dropped power for each curve relative to the maximum dropped power. Therefore, the plot modulation spectra are given by
\begin{equation}
\rho'(\Omega) \equiv \rho(\Omega) \frac{P_{\rm pd0}^2}{P_{\rm pd}^2} = \frac{|\delta \widetilde{P}_{\rm Ts}(\Omega)|^2 / P_s^2}{|\delta \widetilde{P}_{\rm pd}(\Omega)|^2 / P_{\rm pd0}^2}, \label{Def_ModulationSpectra_Plot}
\end{equation}
where $P_{\rm pd0} = 0.85$ mW is the maximum drop power used in Fig.~4d. 

The derivations above take into account only the flapping mechanical mode, since it is most strongly actuated by the optical gradient force. In general, there are many mechanical resonances for the spiderweb resonators, but weakly coupled to the optical waves inside the cavity. In this case, following the same procedure above, it is easy to show that the spectral response of probe modulation now becomes
\begin{equation}
\rho(\Omega) = \left| 2\gamma + \sum_j {{\frac{g_{j}^2}{m_{\rm j} \omega_p {\cal L}_j(\Omega) }} } \right|^2 \frac{P_{\rm pd}^2}{\Gamma_{\rm 0p}^2} \frac{4 \Gamma_{\rm es}^2 \Gamma_{\rm 0s}^2 \Delta_s^2}{\left[ \Delta_s^2 + (\Gamma_{\rm ts}/2)^2 \right]^4}, \label{ProbeModulation_MultiResonance}
\end{equation}
where $g_j$, $m_j$, and ${\cal L}_j(\Omega)$ are optomechanical coupling coefficient, effective motional mass, and the spectral response of mechanical motions, respectively, for the $j^{\rm th}$ mechanical mode. For those weakly actuated mechanical modes, ${\cal L}_j(\Omega) = \Omega_{\rm mj}^2 - \Omega^2 - i \Gamma_{\rm mj} \Omega$ where $\Omega_{\rm mj}$ and $\Gamma_{\rm mj}$ are the resonance frequency and damping rate of the $j^{\rm th}$ mechanical mode. Equation~(\ref{ProbeModulation_MultiResonance}) was used to describe the modulation spectrum shown in Fig.~4(d).

\end{document}